\documentclass[11pt]{article}

\usepackage[T1]{fontenc}
\usepackage[utf8]{inputenc}
\usepackage{lmodern}
\usepackage[margin=1in]{geometry}
\usepackage{amsmath,amssymb,amsfonts}
\usepackage{graphicx}
\usepackage{dcolumn}
\usepackage{bm}
\usepackage{multirow}
\usepackage{array}
\usepackage{booktabs}
\usepackage{tabularx}
\usepackage{xcolor}
\usepackage[numbers,sort&compress]{natbib}
\usepackage[colorlinks=true,linkcolor=blue,citecolor=blue,urlcolor=blue]{hyperref}

\newenvironment{acknowledgments}{\section*{Acknowledgments}}{}
\newcommand{\dodoi}[1]{doi: \href{https://doi.org/#1}{\nolinkurl{#1}}}

\begin{document}
\title{Neural Radiated-Noise Fields for Unmanned Underwater Vehicle Noise Spectrum Prediction in Three-Dimensional Scenes}
\author{Yan Wu \quad Yang Yang \quad Jun Fan \quad Bin Wang\\[0.5em]
\small Key Laboratory of Marine Intelligent Equipment and System, Ministry of Education,\\
\small Shanghai Jiaotong University, Shanghai 200240, PR China\\[0.3em]
\small Corresponding author: \texttt{yang\_dl@sjtu.edu.cn}}
\date{}
\maketitle
\begin{abstract}
Radiated noise in unmanned underwater vehicles (UUVs) is an important indicator for characterizing acoustic signatures and evaluating platform performance. To address the strong dependence of traditional physics-based modeling and numerical simulation methods on target structural information and environmental boundary conditions, and their inability to achieve continuous spatial spectrum-response modeling in three-dimensional scenes, this paper proposes a neural radiated-noise field (NRNF). An NRNF represents the UUV radiated-noise spectrum as a continuous function of the three-dimensional UUV position, the three-dimensional hydrophone position, the UUV yaw angle, and the frequency, enabling query-based prediction at arbitrary spatial locations. The proposed method employs sinusoidal encoding for position and frequency, and introduces a learnable three-dimensional scene feature grid to explicitly represent environmental structure and propagation effects. A spectrum-prediction dataset is constructed from lake trials, and the proposed model is evaluated under three settings: horizontal extrapolation, depth extrapolation, and cross-run generalization. Results show that the NRNF achieves an average prediction error of 3.5 dB in the 50--5000 Hz band. Horizontal extrapolation is easiest, depth extrapolation is the most challenging, and cross-run generalization is of intermediate difficulty. Further ablation results demonstrate that the scene feature grid significantly improves the prediction stability and spatial generalization of the model.  
\end{abstract}

\section{\label{sec:1} Introduction}

Unmanned underwater vehicles (UUVs) have been widely applied in ocean observation, resource exploration, and underwater operations. During navigation, the underwater radiated noise generated by a UUV affects the surrounding acoustic environment and directly determines the detectability of the platform under passive acoustic sensing conditions, as well as its acoustic stealth performance \cite{kita2022passive,mao2024multi,zhang2024experimental}.

The radiated noise spectrum is an important means of analyzing acoustic characteristics and typically comprises both continuous and line spectra \cite{railey2020acoustic}. In shallow-water waveguide environments, the observed spectrum is determined not only by the internal excitation sources of the UUV, the structural transmission paths, and the radiation process, but also by environmental propagation effects, including multipath propagation caused by surface and seabed boundaries, frequency-dependent attenuation and interference, and variations in oceanographic parameters \cite{jensen2011computational}. 

Existing studies on the formation mechanisms and predictive modeling of UUV radiated noise spectra---particularly the line and continuous spectral components---have long relied on theoretical analysis and numerical simulations. These approaches can generally be categorized into several classes: (1) Structural vibration--acoustic radiation coupled modeling methods, which are primarily used to analyze radiated noise generated by structural vibrations of the hull, onboard mechanical equipment, and propulsion systems. Typical approaches include the finite element method \cite{merz2007development}, the boundary element method \cite{kirkup2019boundary}, and hybrid methods combining surface integral formulations with boundary element solvers for exterior acoustic radiation \cite{tomy2024predicting}. (2) Waveguide-based underwater acoustic propagation modeling methods, which describe the propagation and reception of radiated noise in complex ocean environments. Representative techniques include the parabolic equation (PE) method \cite{lin2019three}, normal mode theory \cite{westwood1996normal}, and ray acoustics \cite{d2021bidimensional}. (3) Numerical prediction methods based on flow-induced noise mechanisms, where unsteady flow fields and pressure fluctuations are first computed using computational fluid dynamics, followed by far-field noise prediction using acoustic analogy methods such as the Ffowcs Williams--Hawkings formulation \cite{sezen2022marine,ozden2016underwater}. (4) High-frequency noise analysis methods based on statistical energy and simplified assumptions, such as statistical energy analysis \cite{sheng2004statistical}, are commonly employed for the rapid evaluation of high-frequency vibration and noise energy transmission in complex structural systems. 

Together, these methods constitute an important technical foundation for predicting underwater radiated noise from UUVs and have demonstrated effectiveness across different frequency bands, spatial scales, and application scenarios. Nevertheless, despite their value for physical interpretation and numerical modeling, these approaches generally require relatively detailed information about the target structure and environmental boundary conditions. As a result, they are typically more suitable for high-fidelity modeling in specific scenarios and operating conditions.\par

Replacing traditional numerical solvers with learning-based models has recently led to several novel approaches\cite{mildenhall2021nerf,niemeyer2020differentiable,sitzmann2020implicit}. One class of methods focuses on physics-constrained learning. For example, physics-informed neural networks (PINNs) incorporate the governing acoustic equations directly into the loss function, and the residuals of partial differential equations are computed through automatic differentiation and jointly optimized with boundary and initial conditions \cite{raissi2019physics}. Such models can predict acoustic fields even without explicit training data. Yoon et al. \cite{yoon2024predicting} improved the prediction capability of PINNs for ocean waveguide acoustic fields by introducing a PE envelope representation. To further address challenges such as difficulty learning high-frequency features and unstable training, Tang et al. \cite{tang2025physics} proposed a PINN variant that combines pretraining and adaptive strategies to improve high-frequency representation and generalization. In contrast to PINNs, which rely on physical constraints to compensate for limited data, neural operator learning methods such as DeepONet emphasize learning mappings directly from input function spaces to output fields \cite{lu2021learning}. Once trained, such models can replace repeated numerical iterations with a single forward pass during inference. Xu et al. \cite{xu2023training} proposed using DeepONet as a surrogate solver for the two-dimensional PE model. By using complex acoustic pressure and sound-speed information as inputs, their model learns an approximation to the square-root operator in the PE formulation, enabling more efficient acoustic propagation modeling.\par

Unlike the above approaches, implicit neural representation (INR) typically treats the target response as a queryable function defined over a continuous coordinate domain and directly learns the mapping from input conditions to output responses through coordinate-based neural networks. In recent years, INR has demonstrated strong capability for continuous representation in various acoustic problems. For example, Vengurlekar et al. \cite{vengurlekar2025sh} modeled complex scattering fields using an implicit representation based on spherical harmonic coefficients, learning three-dimensional scattering fields and direction-dependent responses from raw one-dimensional echo signals. In the context of sonar-based three-dimensional reconstruction, Qadri et al. \cite{qadri2022neural} represented scene geometry as a neural implicit function and combined it with differentiable rendering to synthesize sonar observations for dense three-dimensional reconstruction. Notably, INR has also been applied in indoor acoustics to learn continuous mappings from source--receiver position pairs to spatial acoustic responses \cite{ratnarajah2022mesh2ir,su2022inras}. These studies indicate that INR is well suited for describing complex acoustic field responses jointly determined by spatial locations, geometric structures, and propagation conditions.\par

Inspired by these developments, this work introduces INR into the conditional modeling of radiated noise from UUVs and proposes a neural radiated-noise field (NRNF). In this framework, the three-dimensional positions of the UUV and the hydrophone, together with the UUV yaw angle and frequency, are taken as inputs, and the corresponding power spectral density (PSD) is predicted as the output. The model, therefore, learns a continuous spectral-response mapping defined over a three-dimensional acoustic scene. Considering the significant influence of the sea surface, seabed interface, and water medium on acoustic propagation in shallow-water waveguides, the geometric relationship between the UUV and the hydrophone within the entire scene is further encoded explicitly as scene features, which are used as conditional inputs to the NRNF. This design allows the model to capture the influence of environmental structure and propagation effects on PSD formation. Under this framework, the model no longer relies solely on local correspondences between discrete measurement points but instead learns the coupling relationship between scene structure and spectral responses, enabling continuous modeling and prediction of spectral characteristics for previously unseen UUV--hydrophone position pairs.\par

The main contributions of this work can be summarized as follows.
First, unlike traditional underwater radiated-noise modeling approaches that solve the acoustic response for specific operating conditions and receiver locations individually, the proposed NRNF represents the UUV radiated noise PSD as a continuous function defined over three-dimensional space and the frequency domain, enabling unified neural field modeling of spectral responses under varying positions, orientations, and frequencies.
Second, to capture the complex acoustic-field characteristics resulting from the combined effects of propagation paths, boundary reflections, and spatial structures in shallow-water environments, a learnable three-dimensional scene feature grid is introduced to implicitly parameterize the propagation environment. This allows the model to incorporate scene priors during spectral prediction rather than relying solely on explicit coordinate inputs.
Third, the proposed framework jointly considers the directional effects of UUV heading variations on radiated noise and the influence of scene propagation structures on spectral responses, thereby enhancing its ability to represent complex spectral variations. These design choices enable an NRNF to achieve stronger spatial extrapolation capability, improved scene robustness, and more stable prediction performance for radiated noise spectra at previously unseen spatial locations compared with coordinate-based regression models.\par

The remainder of this paper is organized as follows. Section II presents the problem formulation and task definition and introduces the implicit neural representation framework, along with the proposed NRNF model architecture and training objectives. Section III describes the experimental setup and dataset construction process. Section IV presents the prediction results under different test configurations and provides ablation study analyses. Finally, Section V summarizes the paper and discusses potential directions for future work.

\section{\label{sec:2} Model Architecture}

This section aims to develop a model that efficiently predicts the radiated noise spectra of UUVs in a three-dimensional scene under specified conditions. First, a task definition is provided to clarify the problem to be addressed. The core concept of implicit neural networks as a continuous function representation is then briefly reviewed. Building on this foundation, the NRNF is introduced, and its ability to capture UUV radiated-noise spectral characteristics across arbitrary environments is demonstrated. Finally, the parameterization of the NRNF is discussed, enabling the prediction of PSD at arbitrary field points, including those at previously unseen locations.

\subsection{Problem formulation}

The radiated noise of a UUV mainly originates from excitations of the propulsion system and onboard electromechanical equipment, as well as hydrodynamic noise and structural vibrations generated during navigation. These excitations are coupled through the hull structure and radiated into the surrounding water, eventually forming observable noise signals at the receiver. Since the propagation process is strongly influenced by environmental boundary effects and geometric relationships, the spectral characteristics of radiated noise vary with the water environment and the relative source--receiver configuration, including range, depth, and azimuth. In this work, we focus on predicting the UUV radiated-noise PSD in a three-dimensional bounded underwater environment under varying operational conditions and positional configurations.

From a modeling perspective, radiated-noise time-domain signals are typically high-dimensional (with the number of sampling points often reaching the order of $10^{4}$) and exhibit strong randomness. Learning the mapping directly from conditional parameters to time-domain waveforms is therefore computationally expensive and often unstable at capturing key signal characteristics. In contrast, frequency-domain representations more directly reflect the energy distribution of noise across different frequency bands while maintaining a moderate dimensionality, making them more suitable for prediction tasks. 

In particular, when the amplitude spectrum is obtained using a Fourier transform, the spectral estimates in individual frequency bins often exhibit noticeable random fluctuations. The PSD estimation based on Welch's method \cite{welch1967fft} effectively reduces variance through segmentation, windowing, and averaging, producing smoother spectral curves while preserving characteristic spectral lines associated with the target. This property facilitates learning and generalization in neural network models.

A sampled radiated-noise time-domain signal $x(t)$ can be represented as a discrete sequence $x[n]$ under a sampling rate $f_s$. The PSD is estimated using Welch's method as:
\begin{equation}\label{eqn-1}
    \hat{\Phi}_{x} \left ( f \right )  =\frac{1}{M}\sum_{m=1}^{M}\frac{1}{Uf_s}\left | \sum_{n=0}^{L-1} x_{m}\left [ n \right ]w\left [ n \right ]e^{-j2\pi kn/N_{fft} }     \right |^{2}     
\end{equation}
Here, $\hat{\Phi}_{x}(f)$ denotes the PSD estimate of signal $x[n]$ at frequency $f$. The index $m=1,\ldots,M$ represents the segment index, where $M$ is the total number of segments. The term $x_{m}[n]$ denotes the $m$th signal segment of length $L$, and $n=0,\ldots,L-1$ represents the sample index within each segment. The function $w[n]$ denotes the window function, $N_{fft}$ is the FFT length, and $k$ represents the frequency index. The normalization factor of the window function is defined as $U=\frac{1}{L}\sum_{n=0}^{L-1}w^{2}[n]$. For convenience in modeling and evaluation, the PSD is further transformed into a logarithmic scale:

\begin{equation}\label{eqn-2}
  \Phi_{x}(f) = 10\log_{10}\left(\hat{\Phi}_{x}(f)+\varepsilon\right),
\end{equation}
where $\Phi_{x}(f)$ denotes the PSD expressed in decibels and $\varepsilon$ is a small constant introduced for numerical stability.

In this work, the hydrophone is assumed to be omnidirectional, while the UUV radiated noise exhibits directional characteristics. Based on this assumption, the proposed spectral prediction model takes the UUV three-dimensional position $p_u = (x_u, y_u, z_u)$, the hydrophone three-dimensional position $p_h = (x_h, y_h, z_h)$, the UUV yaw angle $\psi$, and the frequency $f$ as inputs, and outputs the corresponding PSD $\Phi_x(f)$.

\subsection{Implicit neural representation }

INR is a class of coordinate-based models designed to represent continuous signals or fields. Unlike traditional approaches that store signals as discrete sampling sequences, INR directly learns a continuous function parameterized by a neural network, enabling signal values to be queried at arbitrary coordinates. 

Let the input space be $X \subseteq \mathbb{R}^{d}$ and the output space be $Y \subseteq \mathbb{R}^{c}$, where $x \in X$ denotes the input coordinate and $y \in Y$ represents the corresponding signal value. The target signal satisfies:

\begin{equation}\label{eqn-3}
u:\mathit{X} \to \mathit{Y} ,u\in \mathit{C} \left ( \mathit{X}, \mathit{Y}\right ),       
\end{equation}
where $\mathit{C}(\mathit{X}, \mathit{Y})$ denotes the set of continuous mappings from $\mathit{X}$ to $\mathit{Y}$. Let $\mathit{K}\subset \mathit{X}$ be a compact set. For any $v\in \mathit{C}(\mathit{K}, \mathit{Y})$, the approximation error is characterized using the uniform norm:

\begin{equation}\label{eqn-4}
 \|v\|_{\infty, K}=\sup _{x \in K}\|v(x)\|        
\end{equation}
Given a finite set of observation samples $\left \{ \left ( x_{i},u\left ( x_{i}  \right )   \right )  \right \} _{i=1}^{N_{s}} $, where $x_{i} \in \mathit{X}$ and $N_{s}$ is the total number of samples, the objective of INR is to learn a parameterized function $\hat{u} \left ( \cdot ;\theta  \right )$ to approximate $u$, such that on $\mathit{K}$:

\begin{equation}\label{eqn-5}
\hat{u}(x;\theta ) \approx u\left ( x \right ) ,x\in \mathit{K}        
\end{equation}
In practice, INR typically adopts a multilayer perceptron (MLP) as a coordinate network to represent the continuous mapping:

\begin{equation}\label{eqn-6}
\hat{u}(x;\theta )=f_{\theta } (x),f_{\theta } :\mathit{X}\to \mathit{Y}     
\end{equation}
where $\theta$ denotes the network parameters. The parameters $\theta$ are learned by minimizing the empirical risk on discrete observation points. Given training samples $\left \{ \left ( x_{i},y_{i} \right ) \right \} _{i=1}^{N_{s}}$, where $y_{i} =u(x_{i})$, the INR optimization objective is formulated as:

\begin{equation} \label{eqn-7}
\theta^{*}=\arg \min _{\theta} \frac{1}{N_{s}} \sum_{i=1}^{N_{s}} \mathcal{L}\left(f_{\theta}\left(x_{i}\right), y_{i}\right)     
\end{equation}
where $\theta^{*}$ denotes the trained network parameters and $\mathcal{L}(\cdot)$ represents the data-fitting loss. By optimizing Eq. (\ref{eqn-7}), the network parameters $\theta^{*}$ implicitly encode the target function, enabling the model not only to fit the sampled observations but also to form a queryable function approximation over the continuous domain.

In practical applications, to better model complex high-frequency structures, Fourier feature mappings are often used to transform the input coordinates into multi-scale Fourier representations. Alternatively, coordinate networks with sinusoidal activation functions can also be employed to improve the representation of high-frequency signal components.

\subsection{Neural Radiated Noise Field (NRNF) }

\begin{figure}[ht]
\centering
\includegraphics[width=0.8\linewidth]{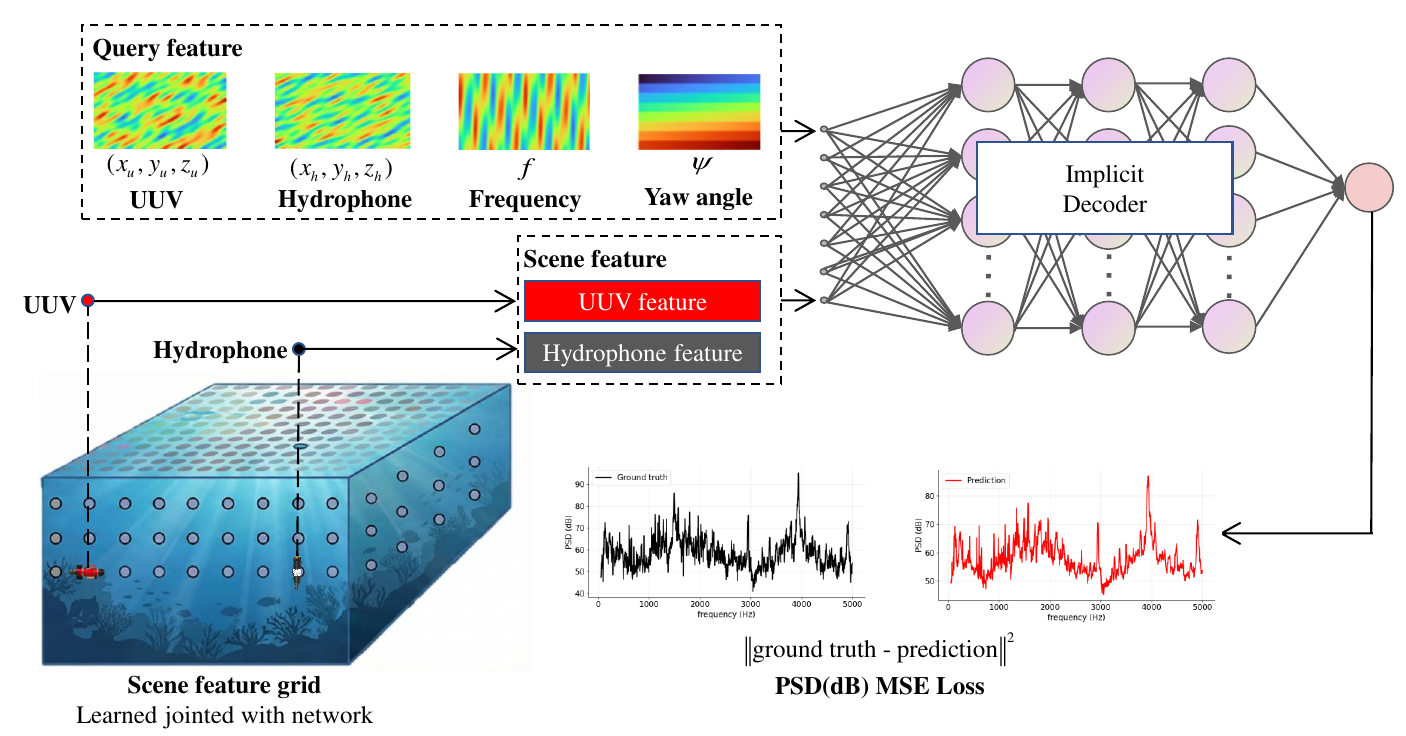}
\caption{\label{fig:FIG1}{ Overall architecture of NRNF. The three-dimensional coordinates of the UUV and hydrophone, yaw angle $\psi$, and frequency $f$ are encoded as query conditions; location-dependent features are sampled from a learnable three-dimensional scene feature grid and fed to an implicit decoder to predict $\Phi _{x}\left ( f \right )  $ at each frequency. The network is trained with an mean squared error loss between predicted and measured PSD.
}}
\end{figure}

An NRNF can be regarded as an instantiation of INR for the task of modeling UUV underwater radiated-noise spectra. It employs an implicit neural network to continuously represent the spectral function, where $\Phi _{x}(f)$ is treated as a continuous mapping of $\left ( p_{u},p_{h},f,\psi    \right ) $ and can be queried at arbitrary spatial coordinates. Unlike conventional INR formulations, an NRNF additionally incorporates learnable environmental feature grids inspired by neural acoustic fields \cite{luo2022learning}. As illustrated in Fig. \ref{fig:FIG1}, these grids allow the model to retrieve environmental context through a cross-attention mechanism, enabling a more effective representation of environmental coupling effects on the radiated-noise PSD. These conditional features are then fed into an implicit decoder, which directly regresses the spectral value $\Phi_x(f)$ at each queried frequency. By performing point-wise prediction across the entire frequency band and concatenating the results, the full PSD curve is obtained:
\begin{align} \label{eqn-8}
\mathrm{NRNF}:\left ( P_{u}, P_{h},f,\psi  \right )  \to \Phi _{x}\left ( f \right )   
\end{align}

It should be emphasized that Eq. (\ref{eqn-8}) does not predict the PSD on a discrete grid. Instead, the spectrum is modeled as a function of continuous variables $(p_u, p_h, \psi, f)$ and queried in a point-wise manner. Since the implicit decoder uses shared parameters, its output is mathematically equivalent to a deterministic continuous mapping. Specifically, for any real-valued spatial coordinates $(p_u, p_h)$ and any frequency $f$, the spectral value $\Phi_x(f)$ can be directly obtained through a forward pass of the network, without relying on training-point indices or additional interpolation procedures. During training, supervision is imposed only at a finite set of observation locations. However, by sharing parameters across samples, the network learns a global functional relationship between coordinates and spectral responses. As a result, the model acquires the ability to query and extrapolate spectral values at previously unseen coordinates. The NRNF architecture consists of two main components: feature composition and implicit decoding. In the feature composition stage, the input parameters are encoded into two types of features: the Query Feature $F_Q$ and the Scene Feature $F_S$. The query feature represents the instantaneous state of the current input parameters, while the scene feature extracts environmental context from the scene feature grid. These features are subsequently fused and fed into the implicit decoder, which maps the combined representation to the spectral value $\Phi_x(f)$. To construct the query feature $F_Q$, sinusoidal positional encoding is applied to the UUV position $p_u$, the hydrophone position $p_h$, and the frequency $f$, enabling the network to better capture high-frequency variations in the spectral function:
\begin{align} \label{eqn-9}
e_{u} =\gamma\left ( p_{u}  \right ),e_{h} =\gamma\left ( p_{h}  \right ),e_{f} =\gamma\left ( f  \right )  
\end{align}
Here, $e_{u}$, $e_{h}$, $e_{f}$ denote the encoded features of $p_{u}$, $p_{h}$, and $f$ respectively. The encoding function $\gamma \left ( \cdot  \right ) $ is constructed using a set of logarithmically spaced frequency bands $\left \{ 2^{0},2^{1} ,...,2^{N_{k} }   \right \} $. For each frequency band, sinusoidal components $\sin \left ( 2\pi \cdot 2^{n_{k} }  \right ) $ and $\cos \left ( 2\pi \cdot 2^{n_{k} }  \right ) $ are computed and concatenated to form the final encoding. Since the yaw angle $\psi$ lies within the range $0^\circ$--$360^\circ$, a learnable embedding is used to encode $\psi$. Specifically, the range is discretized at intervals of $10^\circ$, producing a learnable embedding matrix of dimension $\mathbb{R}^{36\times D}$. Based on the interval in which $\psi$ falls, the corresponding embedding vector in $\mathbb{R}^{1\times D}$ is selected.

To obtain the scene feature $F_S$, the three-dimensional environment is represented as a learnable feature grid. Specifically, the scene is discretized into a set of grid nodes
$P=\{p_i\}_{i=1}^{N_g}$, where $p_i\in\mathbb{R}^3$ denotes the three-dimensional coordinate of the $i$th grid node. Each node is associated with a learnable scene representation $m_i\in\mathbb{R}^{1\times D}$, where $D$ denotes the feature dimension of the grid representation. The scene representation matrix is denoted as $\mathrm{M}=[m_1,\ldots,m_{N_g}]^{\mathsf T}\in\mathbb{R}^{N_g\times D}$, where $N_g$ represents the total number of grid nodes.

For a given query location (e.g., the UUV coordinate $p_u$), a soft-weighted aggregation strategy based on a neighborhood set is adopted to avoid discontinuities introduced by hard nearest-neighbor selection. First, the $K$ nearest grid nodes to $p_u$ are selected from all grid nodes, forming the neighborhood index set
$N_K(p_u)\subset\{1,\ldots,N_g\}$,
where $K$ denotes the number of neighbors. For any $j\in N_K(p_u)$, $p_j$ denotes the coordinate of the neighboring node and $m_j$ represents its corresponding scene token. The distance-based weights are then normalized using a softmax function to obtain $\beta_j(p_u)$:
\begin{equation}\label{eqn-10}
\beta_j(p_u)=
\frac{\exp\!\left(-\|p_u-p_j\|_2^2/\sigma^2\right)}
{\sum\limits_{\ell\in N_K(p_u)}\exp\!\left(-\|p_u-p_\ell\|_2^2/\sigma^2\right)}
\end{equation}
Here, $j,\ell \in N_{K}(p_u)$ are indices of neighboring grid points. The index $j$ denotes the neighbor corresponding to the weight $\beta_j(p_u)$, and $\ell$ is the summation index in the denominator. The scale parameter $\sigma>0$ controls the bandwidth of the Gaussian kernel. 
Using the weights computed from Eq. (\ref{eqn-10}), the neighboring tokens $\{m_j\}$ are aggregated through a weighted combination to construct a location-dependent query token 
$\tilde m(p_u)$:
\begin{equation}\label{eqn-11}
\tilde m(p_u)=\sum_{j\in N_K(p_u)} \beta_j(p_u)\, m_j 
\end{equation}
The resulting $\tilde m(p_u)$ is used as the query, and cross-attention is applied to 
aggregate location-dependent scene context features from the global scene memory $\mathrm{M}$. Specifically, let $W_q$, $W_k$, and $W_v$ denote learnable projection matrices. Let $\alpha_i(p_u)$ represent the relevance weight between the current query token and the $i$th scene token $m_i$. Then
\begin{equation}\label{eqn-12}
\alpha_i(p_u)=
\frac{
\exp\!\left(\dfrac{(W_q\tilde{m}(p_u))^{\top}(W_k m_i)}{\sqrt{D}}\right)
}{
\sum_{n=1}^{N_g}
\exp\!\left(\dfrac{(W_q\tilde{m}(p_u))^{\top}(W_k m_n)}{\sqrt{D}}\right),
}
\end{equation}

\begin{equation}\label{eqn-13}
S(p_u)=\sum_{i=1}^{N_g}\alpha_i(p_u)\,(W_v m_i).
\end{equation}
Here, $\alpha_i(p_u)$ denotes the attention weight associated with the $i$th scene feature, and $S(p_u)$ is the scene-context feature adaptively aggregated at the query location $p_u$. As the query position varies, both $\beta_j(p_u)$ and $\alpha_i(p_u)$ are dynamically updated, which enables scene-conditioned queries at arbitrary continuous spatial coordinates. The hydrophone position $p_h$ is processed in the same manner. According to the reciprocity principle of acoustic propagation, the UUV and hydrophone share the same scene representation $\mathbf{M}$.

The query feature $F_Q$ and scene feature $F_S$ are concatenated along the channel dimension and used as the input to the implicit decoder. The implicit decoder adopts a feedforward MLP architecture composed of eight fully connected layers. Except for the output layer, all layers employ LeakyReLU nonlinear activation functions. To enhance information propagation and training stability, a skip-connection branch is introduced in addition to the main network backbone. This branch consists of two fully connected layers and fuses the mapped output of the input layer with the output of the fourth hidden layer via residual addition. Furthermore, a learnable UUV yaw-angle embedding is incorporated into the intermediate layers. The embedding is injected into the corresponding hidden features through layer-wise additive conditioning. The NRNF model is trained using the mean squared error (MSE) between the predicted and ground-truth spectra. For each training sample, the loss is computed over the set of frequency points $\left\{ f_k \right\}_{k=1}^{N_f}$ as:
\begin{align} \label{eqn-14}
  \mathcal{L}_{\mathrm{NRNF}}=\frac{1}{N_{f}} \sum_{k=1}^{N_{f}}\left\|\Omega\left(p_{u}, p_{h}, \psi, f_{k}\right)-\Phi_{x}\left(f_{k}\right)\right\|_{2}^{2},
\end{align}
where $N_f$ denotes the number of frequency points, and $\Omega(p_u, p_h, \psi, f_k)$ represents the prediction of the NRNF model. The loss $\mathcal{L}_{\mathrm{NRNF}}$ accumulates the prediction error across all frequency points, allowing the model to simultaneously constrain both the global spectral shape and local spectral fluctuations. In this way, the model learns a continuous PSD across different spatial configurations and orientations.

\section{\label{sec:3} Experiments and Dataset}

\subsection{Experimental setup}
\begin{figure}[htp]
\centering
\includegraphics[width=0.98\linewidth]{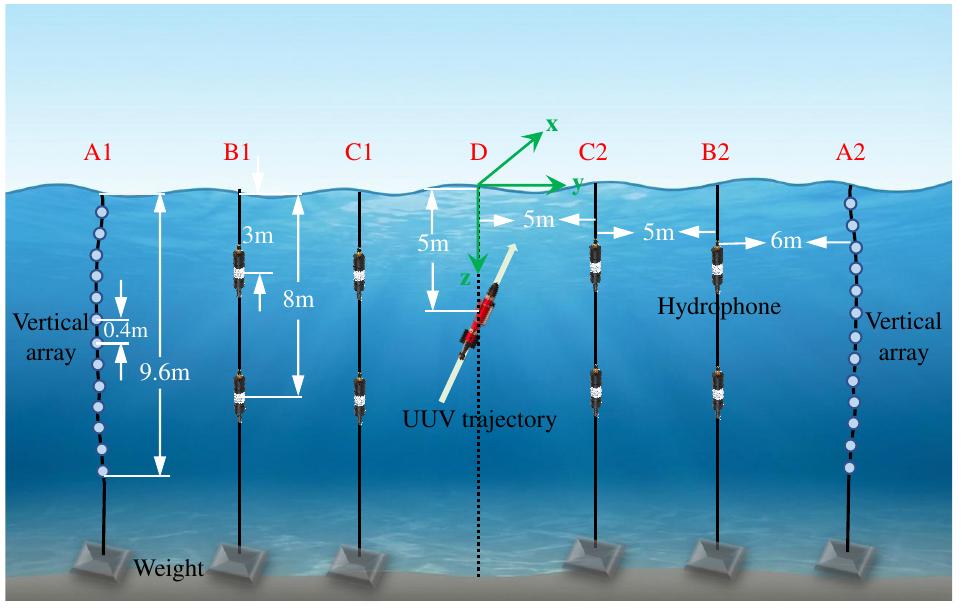}
\caption{\label{fig:FIG2}{Schematic diagram of the experimental setup}}
\end{figure}

The experimental data used in this study were collected in July 2025 at Hekou Reservoir in Huzhou, Zhejiang Province, China. The water depth at the test site was approximately 17 m, and the underwater background noise level was relatively low, providing favorable conditions for underwater acoustic measurements. The experimental setup is illustrated in Fig. \ref{fig:FIG2}.
The measurement system consisted of two symmetrically deployed vertical line arrays and multiple sets of self-contained hydrophones placed at different depths. A1 and A2 denote the two vertical line arrays, each containing 24 array elements with an inter-element spacing of 0.4 m. The first element of each array was positioned 0.1 m below the water surface. In addition, B1 and C1 were located on one side of the central axis D, while B2 and C2 were positioned symmetrically on the opposite side. At location B1, two self-contained hydrophones were suspended by a Kevlar rope attached to a buoy, forming a flexible connection. The deployment depths of the two hydrophones were 3 m and 8 m, respectively. The deployment configurations of B2, C1, and C2 were identical to that of B1. All measurement systems were connected to bottom weights to maintain a vertical orientation and ensure the stability of their relative positions. Geometrically, A1, B1, and C1 were arranged symmetrically with A2, B2, and C2 with respect to the central axis D. Their horizontal distances from the axis were 16 m, 10 m, and 5 m, respectively. The UUV was operated at a navigation depth of 5 m and traveled along a trajectory within the vertical measurement plane, crossing the central axis D from the far side to the near side. A single run was defined as one complete trajectory in which the UUV traveled from the far side to the near side and then back again. Data were recorded from four runs at each propulsion speed of 2000, 3000, and 4000 revolutions per minute (RPM).

The UUV platform and its propeller were made of aluminum alloy. 
The key structural characteristics and operating parameters are summarized in Table~\ref{tab1:essential}. The UUV had a total length of 1.121 m and a main body diameter of 0.124 m. The propeller model used in the experiment was the RHINCODON-ME207. During the experiment, the relative position between the measurement system and the UUV was determined using a master--slave acoustic beacon system. A master beacon was deployed at the location of the A1 vertical array and periodically transmitted interrogation signals. A slave beacon mounted on the UUV returned a response signal upon receiving the interrogation. The distance between the beacons was calculated from the round-trip propagation time of the acoustic signal, thereby providing the relative geometric relationship between the UUV and the measurement array. Meanwhile, the onboard inertial navigation system of the UUV recorded state parameters in real time, including the depth, yaw angle, pitch angle, roll angle, and rotational speed. These measurements provided synchronized operational parameters for subsequent spectral modeling and conditional input construction.

To establish a noise baseline and evaluate the incremental radiated noise generated during UUV operation, ambient underwater noise was first recorded at the beginning of the experiment. These background measurements were later used for comparative analysis with the noise data measured during UUV operation.

\begin{table}[ht]
\centering
\caption{\label{tab1:essential} Parameters of the UUV and propeller}
\begin{tabular}{@{}lll@{}}
\toprule
Category & Essential factor & Value \\
\midrule
\multirow{3}{*}{UUV}
  & Length             & 1.121\,m \\
  & Principal diameter & 0.124\,m \\
  & Working depth      & 0--200\,m \\
\midrule
\multirow{3}{*}{Propeller}
  & Diameter           & 0.99\,m \\
  & Blade number       & 3 \\
  & Design speed       & 3200\,RPM \\
\bottomrule
\end{tabular}
\end{table}

\subsection{Dataset construction}

This section describes the processing of the radiated-noise signals recorded by the hydrophones and the construction of the dataset used for model training. First, a three-dimensional Cartesian coordinate system was established, as illustrated in Fig. \ref{fig:FIG2}. The origin was defined at point D. The direction from D toward A2 was defined as the positive $y$-axis, while the downward underwater direction was defined as the positive $z$-axis. The positive $x$-axis was then determined according to the right-hand rule. The UUV position information from the inertial navigation system was transformed into this coordinate system, and the hydrophone positions were converted to the same coordinate system using deployment geometry and localization information, ensuring geometric consistency in subsequent modeling.

For signal processing, each sample segment was defined as a 1.2 s time-domain signal. All raw acoustic signals were first resampled to a unified sampling rate of $f_s=32\,\mathrm{kHz}$, after which the PSD was estimated using Welch's method. The Welch parameters were configured as follows: the segment length was $L_{seg}=0.4\,\mathrm{s}$, the segment overlap ratio was 0.75, and the number of FFT points was set to $N_{fft}=f_s\cdot L_{seg}$. Under this configuration, each PSD curve contained 1981 frequency bins, corresponding to $\Phi_x(f)\in\mathbb{R}^{1981\times1}$, with the frequency vector $f\in\mathbb{R}^{1981\times1}$.

\begin{figure}[htp]
\centering
\includegraphics[width=0.9\linewidth]{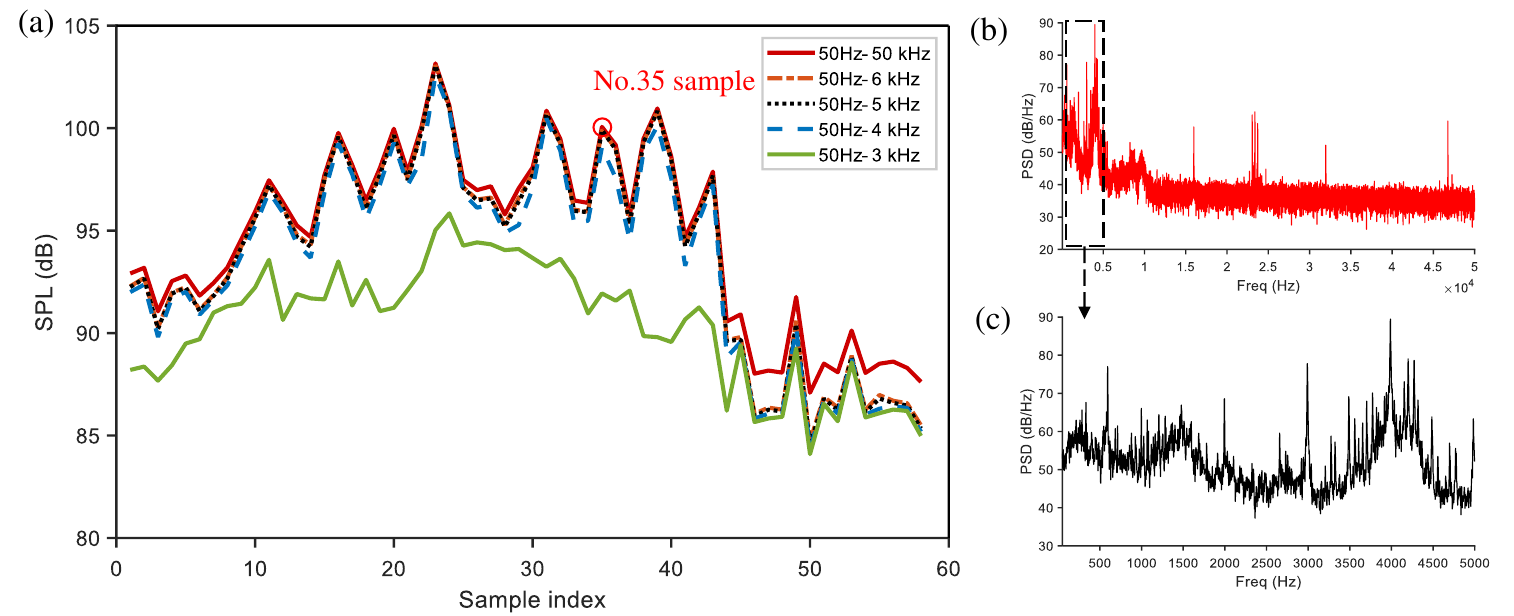}
\caption{\label{fig:FIG3}{Example of UUV radiated-noise sound pressure level and power spectral density. (a) Variation of the SPL of UUV radiated noise with respect to the sample index within one run. Different curves correspond to SPL values calculated over different frequency bands. (b) and (c) show the PSD of the No. 35 sample.  (b) presents the PSD over 50--50000 Hz, while  (c) shows the PSD over 50--5000 Hz. It can be observed that most tonal components and dominant energy of the radiated noise are concentrated in the 50--5000 Hz band.}}
\end{figure}

To determine the prediction frequency band of the proposed model, the in-band sound pressure level (SPL) and spectral structure of the radiated noise within a single run were statistically analyzed. The results are shown in Fig. \ref{fig:FIG3}. Figure \ref{fig:FIG3}(a) presents the variation of the SPL with respect to the sample index within a run, where the in-band SPL was computed for five frequency bands: 50--3000 Hz, 50--4000 Hz, 50--5000 Hz, 50--6000 Hz, and 50--50000 Hz. It can be observed that the in-band SPL values for 50--4000 Hz, 50--5000 Hz, and 50--6000 Hz closely match the broadband SPL over 50--50000 Hz, with differences within approximately 0.5 dB. This indicates that the broadband energy is primarily contributed by frequencies below 6000 Hz.

Further details are illustrated in Fig. \ref{fig:FIG3}(b) and Fig. \ref{fig:FIG3}(c), which show PSD examples for sample No. 35. Figure \ref{fig:FIG3}(b) presents the full-band spectrum from 50 Hz to 50000 Hz, while Fig. \ref{fig:FIG3}(c) shows a refined view of the spectrum from 50 Hz to 5000 Hz. As seen in Fig. \ref{fig:FIG3}(c), the dominant narrowband line spectra and significant energy fluctuations generated by the propulsion system and electromechanical components are mainly concentrated within the 50--5000 Hz range. In contrast, frequencies above 5000 Hz in Fig. \ref{fig:FIG3}(b) primarily exhibit relatively weak high-frequency broadband components, which provide limited additional information for target characterization and operational-state representation. Based on the above observations, the prediction frequency band of the proposed model was selected as 50--5000 Hz, in order to capture the main spectral characteristics while maintaining a manageable output dimensionality and model complexity. In addition, several clusters of narrowband spectral lines appearing beyond 10 kHz in Fig. \ref{fig:FIG3}(b) can be attributed to sideband clusters generated by pulse-width modulation switching frequencies interacting with low-frequency spectral components.

Each sample in the dataset consists of two parts: the condition and the observation. The observation corresponds to the radiated-noise PSD $\Phi_x(f)$ of the sample together with its frequency vector $f$. The condition describes the spatial and orientation parameters under which the PSD is generated. Specifically, it includes the UUV position in the constructed Cartesian coordinate system $\left(x_u, y_u, z_u\right)$ at the sampling instant, the position of the receiving hydrophone $\left(x_h, y_h, z_h\right)$ in the same coordinate system, and the yaw angle $\psi$ of the UUV. Accordingly, the dataset can be represented as a collection of $N_s$ samples $\mathcal{D}=\{(c_i, o_i)\}_{i=1}^{N_s}$, where $c_i$ denotes the condition parameters of the $i$th sample and $o_i$ denotes the corresponding observation.

\begin{subequations}\label{eqn-15}
\begin{align}
\mathcal{D} &= \left\{\left(\mathbf{c}_n,\mathbf{o}_n\right)\right\}_{n=1}^{N_s}, \\
\mathbf{c}_n &= (x_u,y_u,z_u,x_h,y_h,z_h,\psi), \\
\mathbf{o}_n &= \left(\Phi_x(f),\,f\right),
\end{align}
\end{subequations}
where $\mathbf{c}_n$ and $\mathbf{o}_n$ denote the condition and observation of the $n$th sample, respectively. 
The dimensionality and value ranges of the sample parameters are summarized in Table \ref{tab2:sample_params_ranges}. The experimental acquisition system consisted of two 24-element vertical line arrays and eight self-contained hydrophones, yielding a total of 56 receiving channels. Under the three UUV rotational-speed operating conditions, a total of 44820 samples were collected, corresponding to an overall recording duration of approximately 14.94 hours.

\begin{table}[ht]
\centering
\caption{\label{tab2:sample_params_ranges} Sample parameters and data ranges}
\small
\begin{tabularx}{\textwidth}{@{}l>{\raggedright\arraybackslash}Xc>{\raggedright\arraybackslash}p{0.28\textwidth}@{}}
\toprule
Symbol & Description & Dimension & Numerical value \\
\midrule
$x_u$ (m)       & UUV x-coordinate        & $\mathbb{R}^{1}$    & $[-105,\,95]$ \\
$y_u$ (m)       & UUV y-coordinate        & $\mathbb{R}^{1}$    & $[-21,\,23.5]$ \\
$z_u$ (m)       & UUV z-coordinate        & $\mathbb{R}^{1}$    & $[2,\,5]$ \\
$x_h$ (m)       & Hydrophone x-coordinate & $\mathbb{R}^{1}$    & $0$ \\
$y_h$ (m)       & Hydrophone y-coordinate & $\mathbb{R}^{1}$    & $[-16,\,-10,\,-5,\,5,\,10,\,16]$ \\
$z_h$ (m)       & Hydrophone z-coordinate & $\mathbb{R}^{1}$    & $[0.1:0.4:9.3,\ 3,\ 8]$ \\
$\psi$ (rad)    & UUV yaw angle           & $\mathbb{R}^{1}$    & $[-2.8449,\,3.1067]$ \\
$\mathrm{PSD}_{\mathrm{dB}}$ (dB) & Welch PSD in dB & $\mathbb{R}^{1981}$ & $[81,\,105]$ \\
$f$ (Hz)        & Frequency vector        & $\mathbb{R}^{1981}$ & $[50:2.5:5000]$ \\
\bottomrule
\end{tabularx}
\end{table}

To systematically evaluate the generalization capability and robustness of the proposed model under different observation conditions, three experimental configurations, denoted as Settings I--III, were designed. The corresponding training and testing splits are summarized in Table \ref{tab3:settings_splits}. Setting I was designed to examine the model's horizontal transfer capability. Within the same run, the receiver data located on the negative side of the $y$-axis (A1, B1, C1) were used as the training set, while the receiver data on the positive side of the $y$-axis (A2, B2, C2) were used as the test set. This configuration evaluated the model's extrapolation performance with respect to changes in horizontal receiver positions. Setting II evaluated the model's extrapolation capability along the depth dimension. The training set consisted of receiver data at depths shallower than 5 m, including the shallow array elements of the two vertical line arrays and the self-contained hydrophones deployed at 3 m depth. The test set consisted of receiver data deeper than 5 m, including the deeper array elements of the two vertical line arrays and the self-contained hydrophones deployed at 8 m depth. It should be noted that, for Run 4, all elements of the A2 array $(0.1:0.4:9.3)\,\mathrm{m}$ were used for testing to ensure that the evaluation in this run covered the complete receiver depth range. Setting III evaluated cross-run generalization and assessed the robustness of the model with respect to environmental perturbations and variations in UUV trajectory execution. The training set consisted of data from all hydrophones in Runs 1--3, while the test set consisted of the corresponding hydrophone data from Run 4.

\begin{table}[ht]
\centering
\caption{\label{tab3:settings_splits} Three test settings and corresponding training/test splits.}
\footnotesize
\renewcommand{\arraystretch}{1.18}
\begin{tabularx}{\textwidth}{@{}c>{\raggedright\arraybackslash}p{0.19\textwidth}>{\raggedright\arraybackslash}X>{\raggedright\arraybackslash}X@{}}
\toprule
Setting & Test objective & Training set & Test set \\
\midrule
I &
Horizontal extrapolation capability &
Runs 1--4\newline A1, B1, C1 &
Runs 1--4\newline A2, B2, C2 \\
\midrule
II &
Depth extrapolation capability &
Runs 1--4\newline A1, A2: $0.1{:}0.4{:}4.9$ m\newline B1, C1, B2, C2: 3 m &
Runs 1--4\newline A1: $5.3{:}0.4{:}9.3$ m\newline B1, C1, B2, C2: 8 m\newline A2: $5.3{:}0.4{:}9.3$ m$^{\ast}$ \\
\midrule
III &
Cross-run generalization capability &
Runs 1--3\newline A1, B1, C1\newline A2, B2, C2 &
Run 4\newline A1, B1, C1\newline A2, B2, C2 \\
\bottomrule
\end{tabularx}
\vspace{2pt}
\begin{flushleft}
\footnotesize $^{\ast}$~Erratum: In Run 4, A2 uses all elements $0.1{:}0.4{:}9.3$ m for testing.
\end{flushleft}
\end{table}

\section{Results and Discussion}
This section evaluates the proposed NRNF model from several perspectives, 
including prediction performance, generalization capability, and effectiveness analysis. In the implementation of the NRNF, the query feature $F_Q$ was first normalized such that the three-dimensional positions of the UUV and hydrophone, as well as the frequency variable, were linearly scaled to the interval $\left[-1,1\right]$. The spatial coordinates $p_u$ and $p_h$ were encoded using Fourier feature mappings, with the highest frequency scale set to $2^{7}$. The frequency variable $f$ was encoded in the same manner, with the highest frequency scale set to $2^{10}$. To explicitly model the directional characteristics of the radiated noise, a learnable yaw-angle embedding matrix $\mathrm{E}\in\mathbb{R}^{7\times36\times512}$ 
was introduced for the UUV orientation, where $7$ denotes the number of intermediate network layers that receive the embedding, $36$ denotes the number of discretized azimuth bins, and $512$ denotes the feature dimension. In addition, to capture environmental information and spatial priors, a learnable three-dimensional scene feature grid was constructed. The computational domain was defined as the three-dimensional space 
$x\in\left[-110,100\right],\;y\in\left[-25,25\right],\;z\in\left[0,10\right]$. The grid resolutions were set to $\triangle x=2$, $\triangle y=1$, and $\triangle z=1$, resulting in a total of $59466$ grid nodes. Each grid node was assigned a 64-dimensional learnable feature vector initialized from an independent Gaussian distribution $\mathcal{N}\!\left(0,\frac{1}{\sqrt{64}}\right)$. In the cross-attention module for scene--grid interaction, the feature dimension was set to 64, and the number of attention heads was set to 4. Furthermore, to enhance the attention mechanism's ability to capture frequency-dependent variations, a frequency encoding with six components was introduced into the attention module. The implicit decoder network used a hidden feature width of 512 and LeakyReLU activation with a negative slope of 0.1. 

The model was trained using the Adam optimizer \cite{kingma2014adam}. 
Both the network parameters and the scene--grid parameters shared an initial learning rate of $5\times10^{-4}$. The learning rate scheduler used a decay factor of 0.5 with a patience of 15 epochs, and the minimum learning rate was set to $10^{-6}$. All experiments were conducted on an NVIDIA GeForce RTX 4090 GPU with CUDA 12.6 for accelerated computation. To quantitatively evaluate the model performance, the root mean squared error (RMSE) and mean absolute error (MAE) were adopted as evaluation metrics:
\begin{align} 
 R M S E=\sqrt{\frac{1}{N_f} \sum_{i=1}^{N_f}\left(y_{i}-\hat{y}_{i}\right)^{2}},
\end{align}
\begin{align} 
 M A E=\frac{1}{N_f}\sum_{i=1}^{N_f}\left | y_{i}-\hat{y}_{i}    \right |,   
\end{align}
where $y_i$ denotes the measured value of the $i$th PSD frequency bin under a given test condition, and $\hat{y}_i$ denotes the corresponding predicted value. $N_f$ represents the total number of frequency bins in the PSD, with $i=1,2,\ldots,N_f$.

\subsection{Horizontal extrapolation validation}

To evaluate the NRNF prediction performance under horizontal variations of receiver positions, Setting I defined in Table~\ref{tab3:settings_splits} was adopted for validation. In this setting, the training set comprised the receiver data from A1, B1, and C1, while the test set comprised the receiver data from A2, B2, and C2. 
The primary difference between the two sets lay in the $y$-axis positions of the receiving hydrophones, while the depth range of the receivers remained the same. Therefore, this setting was mainly used to examine the model's extrapolation capability when horizontal receiver positions change while the depth distribution remains unchanged.

\begin{figure}[htp]
\centering
\includegraphics[width=0.98\linewidth]{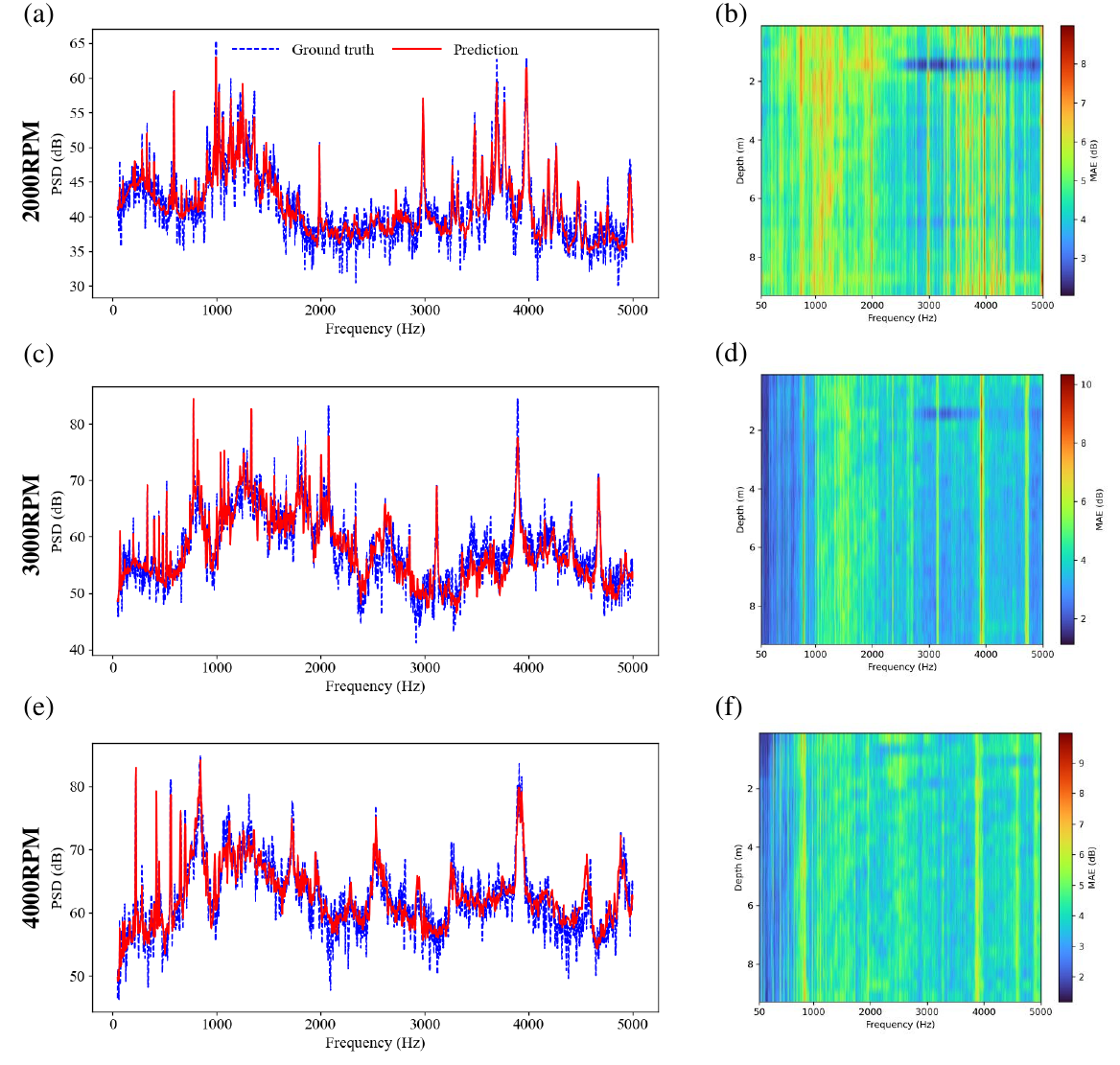}
\caption{\label{fig:FIG4}{PSD prediction under Setting I. Panels (a), (c), and (e) show representative comparisons between measured and predicted spectra at shaft rotational speeds of 2000, 3000, and 4000 RPM, respectively, for a receiver located approximately 50 m from the UUV. Panels (b), (d), and (f) present the corresponding depth--frequency distributions of the MAE for 2000, 3000, and 4000 RPM.}}
\end{figure}

Figures~\ref{fig:FIG4}(a), (c), and (e) present the PSD prediction results under three propulsion speeds: RPM = 2000, 3000, and 4000. The corresponding UUV positions $\left[x,y,z\right]$ are $\left[-48.2,1.5,5.0\right]$, $\left[-49.5,0.7,5.0\right]$, and $\left[-48.7,0.2,5.0\right]$, respectively. The selected receiver is the 13th element of the A2 vertical array, located at $\left[0.0,-16.0,4.9\right]$. In this example, the distance between the UUV and the hydrophone was approximately 50 m, which can be regarded as a representative case under mid-to-far-field reception conditions. The corresponding prediction errors were $\left[\mathrm{RMSE},\mathrm{MAE}\right]_{\mathrm{RPM}=2000}=\left[3.8,3.0\right]$, 
$\left[\mathrm{RMSE},\mathrm{MAE}\right]_{\mathrm{RPM}=3000}=\left[3.0,2.3\right]$, and $\left[\mathrm{RMSE},\mathrm{MAE}\right]_{\mathrm{RPM}=4000}=\left[3.2,2.6\right]$. The detailed numerical results are summarized in Table~\ref{tab:rmse_mae}. It can be observed that, under all three propulsion speeds, the RMSE values remained below 4 dB and the MAE values were approximately within 2--3 dB. These results indicate that the NRNF maintains good prediction accuracy under horizontal variations of receiver positions. Among the three operating conditions, the RPM = 2000 case exhibits slightly larger errors, which may be attributed to more complex spectral fluctuations and more pronounced variations in local frequency components.

Figures~\ref{fig:FIG4}(b), (d), and (f) show the distribution of the MAE 
with respect to the receiver depth and frequency. Since the variation trends of the RMSE and MAE are generally consistent, only the MAE is presented here for visualization. It can be observed that the variation of the error along the depth dimension is relatively small. This is mainly because the training data already covers the entire receiver depth range, so the model does not encounter significant extrapolation difficulty along the depth dimension in this setting. In contrast, the prediction error is mainly concentrated in several specific frequency bands, particularly around 1000 Hz, 2000 Hz, 3000 Hz, and 4000 Hz. These frequency bands typically correspond to regions where spectral lines are dense or where local spectral fluctuations are more pronounced. The complex spectral structures in these regions impose higher requirements on the model's spectral representation and fitting capability. 
Overall, the NRNF maintains stable spectral prediction performance 
under horizontal spatial extrapolation conditions. The primary source of the prediction error arises from the difficulty of fitting complex spectral structures rather than from horizontal changes in receiver positions.

\begin{table}[ht]
\centering
\caption{\label{tab:rmse_mae} RMSE and MAE under three test settings at different rotational speeds.}
\begin{tabular}{@{}llcc@{}}
\toprule
Setting & Rotational speed & RMSE/dB & MAE/dB \\
\midrule
\multirow{3}{*}{Setting I} 
& 2000 RPM & 3.8 & 3.0 \\
& 3000 RPM & 3.0 & 2.3 \\
& 4000 RPM & 3.2 & 2.6 \\
\midrule
\multirow{3}{*}{Setting II} 
& 2000 RPM & 5.2 & 4.0 \\
& 3000 RPM & 5.1 & 3.9 \\
& 4000 RPM & 5.9 & 4.4 \\
\midrule
\multirow{3}{*}{Setting III} 
& 2000 RPM & 4.9 & 3.8 \\
& 3000 RPM & 4.8 & 3.6 \\
& 4000 RPM & 4.9 & 3.7 \\
\bottomrule
\end{tabular}
\end{table}

\subsection{Depth-wise extrapolation validation}

\begin{figure}[htp]
\centering
\includegraphics[width=0.98\linewidth]{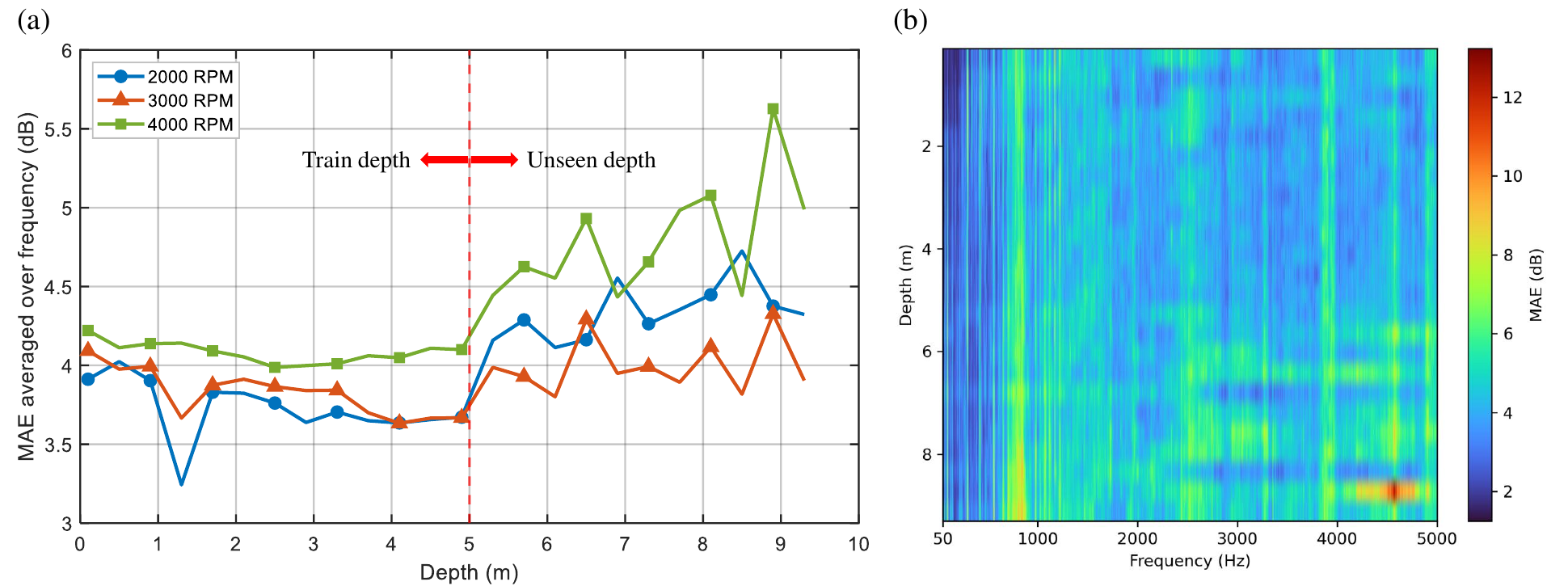}
\caption{\label{fig:FIG5}{Prediction errors under Setting II. (a) MAE averaged over frequency, as a function of receiver depth for propulsion speeds of 2000, 3000, and 4000 rpm. The dashed vertical line at 5 m denotes the boundary between the training depth range ($> 5m$) and the unseen depth range ($\le 5m$).(b) Depth--frequency MAE distribution at 4000 rpm.}}
\end{figure}

Figure~\ref{fig:FIG5} presents the training and testing results of the NRNF under Setting II. This setting is designed to evaluate the model's generalization performance when the receiver depth varies. The training samples were obtained from shallow-depth receiver data, while the test samples primarily came from deeper-water layers not included in the training set. In particular, for Run 4, the A2 vertical array used the full-depth data for testing. Table~\ref{tab:rmse_mae} summarizes the error statistics of Setting II under different propulsion speeds. 
Compared with Setting I, the prediction errors increase noticeably under all three operating conditions. The RMSE values all exceed 5 dB, and the MAE values approach or exceed 4 dB. The average error across the three operating conditions is $\left[\overline{\mathrm{RMSE}},\overline{\mathrm{MAE}}\right]=\left[5.4\,\mathrm{dB},4.1\,\mathrm{dB}\right]$. These results indicate that the prediction task becomes significantly more challenging when the test depths fall outside the depth range covered by the training data. The average MAE curves along the receiver depth for different propulsion speeds are shown in Fig. \ref{fig:FIG5}(a). The region on the right side of the figure corresponds to the unseen depth interval that is not included in the training data. It can be observed that, within the training depth range, the MAE values remain relatively stable for all operating conditions. 
However, once entering the unseen depth region, the errors exhibit an overall increasing trend and show more pronounced fluctuations at larger depths. Within the depth range of 5--9.3 m, the average MAE increases by approximately 0.6 dB, 0.2 dB, and 0.7 dB compared with the 0.1--5 m range under 2000, 3000, and 4000 RPM conditions, respectively. These results indicate that although the model maintains stable prediction accuracy within the observed depth range, the prediction error gradually accumulates as the receiver depth moves beyond the training coverage.

Figure \ref{fig:FIG5}(b) further shows the two-dimensional distribution of the MAE with respect to receiver depth and frequency under the 4000 RPM condition. The errors are generally higher in the unseen depth region, with localized error amplification in several specific frequency bands. This phenomenon is particularly evident in regions where the spectral structure exhibits stronger fluctuations. 
These observations suggest that variations in receiver depth not only alter the acoustic propagation paths and interference structures of the sound field, but also increase the difficulty of fitting complex spectral regions. Consequently, the model faces greater prediction challenges under depth-extrapolation conditions.

\subsection{Cross-run generalization validation}

Both Setting I and Setting II involved training--testing splits based on data from Runs 1--4. The main difference between the two settings lies in the way the spatial dimension was partitioned. Under Setting I, the model's transfer capability was evaluated with respect to horizontal variations of receiver positions, whereas under Setting II, the model's extrapolation capability was examined along the receiver depth dimension. In both settings, the training and testing samples were drawn from the same set of runs. Therefore, the environmental conditions, operating states, and experimental configurations 
remained statistically consistent between the training and test data. In contrast, Setting III employed a cross-run data partitioning strategy. The training data were obtained from Runs 1--3, while the test data came from Run 4, which was not involved in the training process. Since different runs may involve subtle variations in environmental parameters, operational states, and experimental conditions, Setting III introduced a more pronounced distribution shift in the data. This configuration, therefore, provides a suitable scenario for evaluating the generalization robustness of the model under distribution shift conditions.

\begin{figure}[htp]
\centering
\includegraphics[width=0.8\linewidth]{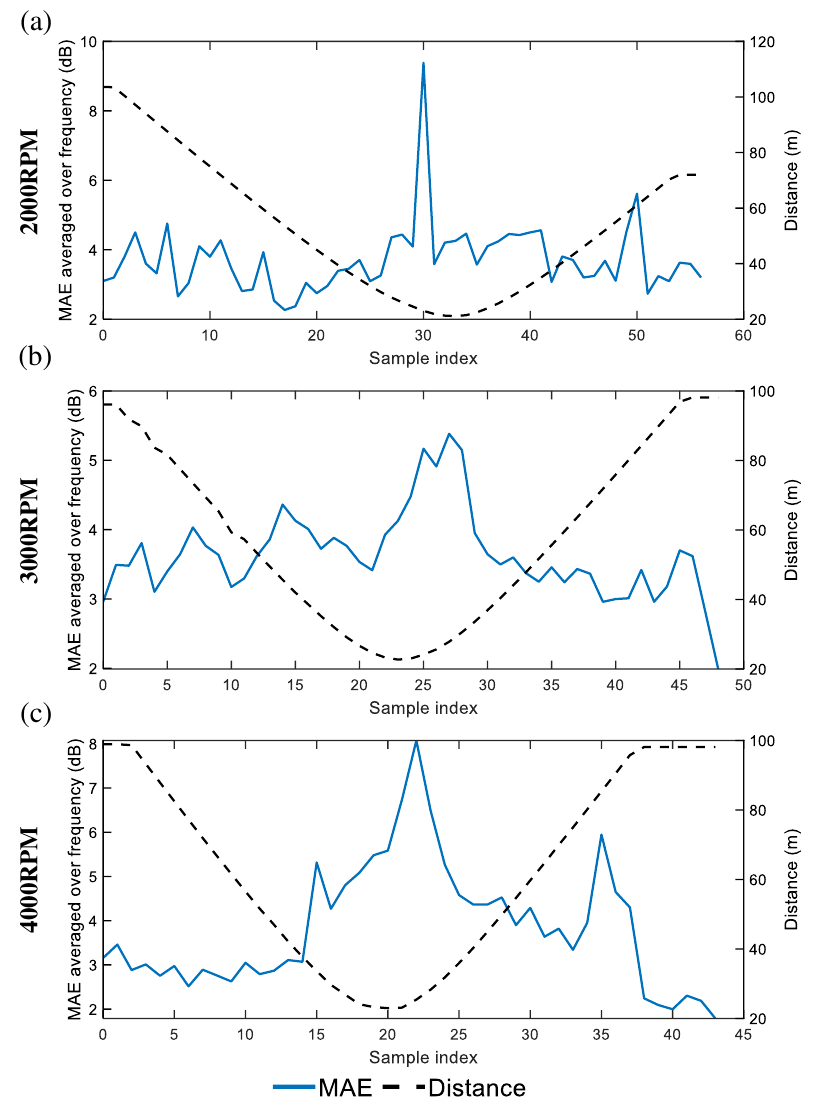}
\caption{\label{fig:FIG6}{Prediction errors under Setting III. Frequency-averaged MAE (blue, left axis) and corresponding UUV--receiver distance (black dashed, right axis) as functions of track sample index for (a) 2000, (b) 3000, and (c) 4000 rpm.}}
\end{figure}

Table \ref{tab:rmse_mae} summarizes the error statistics for Setting III. The average error across the three operating conditions is 
$\left[\overline{\mathrm{RMSE}},\overline{\mathrm{MAE}}\right]=\left[4.9\,\mathrm{dB},3.7\,\mathrm{dB}\right]$. Compared with Setting I, the error level in Setting III is slightly larger. However, the overall error remains lower than that observed in Setting II. This indicates that although cross-run partitioning introduces a certain degree of distribution shift, 
its impact is still smaller than the sound-field structural variations caused by depth extrapolation. Figures~\ref{fig:FIG6}(a)--(c) show the frequency-averaged MAE along the sampling sequence of each run under the three propulsion-speed conditions, together with the corresponding UUV--hydrophone distance. Within each run, the distance follows a characteristic pattern of ``far-to-near, then near-to-far''. Correspondingly, the MAE remains relatively low during the far-range stage but increases noticeably 
as the UUV approaches the hydrophone. After the distance increases again, the error level gradually decreases. This trend is consistent across all three propulsion speeds, indicating that the additional errors in Setting III are mainly concentrated during the near-range reception stage. This observation suggests that spectral prediction becomes more challenging under short-range conditions. On the one hand, the received signal is more susceptible to multipath interference from reflections off the sea surface and seabed. On the other hand, as the propagation path shortens, frequency-dependent propagation effects and local structural variations exert a stronger influence on the spectral shape. As a result, the PSD exhibits higher line density and stronger local fluctuations, 
leading to a more complex spectral structure. These factors collectively increase the difficulty of model fitting, thereby causing larger prediction errors in the near-range region.

\subsection{Scene feature conditioning and its role in generalization}

When using an NRNF to predict the PSD of UUV radiated noise, the key lies in effectively parameterizing the underlying acoustic field. The model should not only fit the spectral responses corresponding to the UUV--hydrophone position pairs observed during training, but also provide reliable predictions for unseen position pairs during testing. If the mapping in Eq. (\ref{eqn-8}) is directly parameterized and learned, the network may simply treat the coordinates of the UUV and hydrophone as 
a combination of input variables. This may lead to overfitting and weaken the model's ability to distinguish the physical roles of the source and receiver. In practice, the propagation of UUV radiated noise from the source to the receiving point depends not only on their relative positions but also on environmental factors such as surface and seabed boundaries and medium attenuation. Therefore, to improve the model's generalization capability for unseen position pairs, relying 
solely on explicit coordinate inputs is often insufficient. It is also necessary to capture the geometric relationships between the UUV and the hydrophone in the propagation environment, along with the implicit structural information of the acoustic field.

\begin{figure}[htp]
\centering
\includegraphics[width=0.9\linewidth]{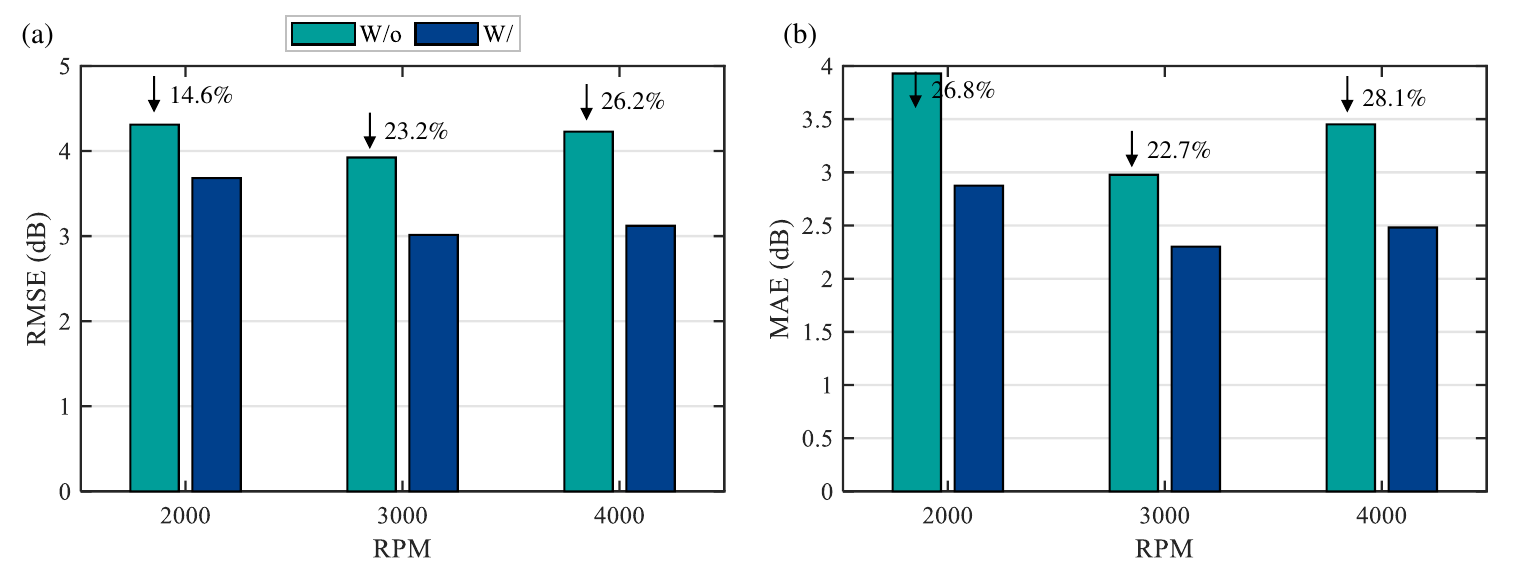}
\caption{\label{fig:FIG7}{(a) RMSE and (b) MAE comparisons between the model without (W/o) and with (W/) global scene-grid features for 2000, 3000, and 4000 rpm. Numbers above bars denote the relative error reduction achieved by introducing the global grid features.}}
\end{figure}

Motivated by these considerations, a learnable scene feature grid is introduced within the three-dimensional computational domain, as illustrated in Fig.~\ref{fig:FIG1}, to enhance the model's representation of the propagation environment. The grid is discretized throughout the entire water domain and is jointly optimized together with the network parameters. During prediction, the model queries the scene-grid features according to the positions of the UUV and hydrophone, and the resulting scene context is provided to the implicit decoder as additional conditional input. In this way, the prediction depends not only on explicit coordinate encoding but also on the latent propagation-structure priors embedded in the environment. To verify the effectiveness of this design, we further compare two model variants: one without the scene feature grid (W/o) and the other with it (W/). As shown in Fig.~\ref{fig:FIG7}, introducing the scene feature grid significantly reduces both the RMSE and MAE under all three propulsion-speed conditions, with improvements of approximately 14\%--28\%. Notably, this reduction 
in error is not limited to a specific operating condition or a single evaluation metric. Instead, consistent improvements are observed across different propulsion speeds and error measures. These results demonstrate that the propagation-environment priors provided by the scene feature grid effectively enhance the model's ability to represent and predict spectral responses for unseen position pairs. Consequently, conditioning the NRNF model on scene structure is not merely an optional design component but rather a key element for achieving stable generalization, improved spatial extrapolation, and enhanced robustness.

\section{\label{sec:4} CONCLUSIONS}

This paper has proposed an NRNF model to predict the PSD of UUV radiated noise in three-dimensional environments. The proposed approach represents the PSD of UUV radiated noise as a continuous mapping with respect to the three-dimensional position of the UUV, the three-dimensional position of the hydrophone, the UUV yaw angle, and the frequency. This formulation enables spectral-response queries at arbitrary spatial coordinates, including position pairs that are not observed during the training stage. To enhance the model's capability to represent directional characteristics and propagation-environment structures, the NRNF employs sinusoidal positional encoding for spatial coordinates and frequency, introduces a learnable discrete embedding for the yaw angle, and further constructs a learnable three-dimensional scene feature grid. Through an attention mechanism, the model retrieves scene-context information from this grid, thereby explicitly modeling environmental structure and acoustic propagation effects.

Based on lake experiment measurements, an evaluation framework for UUV radiated-noise spectrum prediction was established, together with three progressive testing settings corresponding to horizontal receiver-position transfer, depth extrapolation, and cross-run generalization. Experimental results demonstrate that the NRNF achieves an average prediction error of approximately 3.5 dB within the 50--5000 Hz frequency band. Moreover, the model exhibits consistent generalization patterns across different testing settings: horizontal spatial transfer is the easiest scenario, depth extrapolation is the most challenging, and cross-run generalization lies 
between the two. These observations indicate that, compared with horizontal 
position changes and cross-run distribution shifts, the vertical propagation structure variations caused by receiver-depth changes have a more significant impact on spectral prediction. Further ablation experiments showed that incorporating the scene feature grid consistently reduces prediction errors across all operating conditions. This result confirms that explicitly conditioning the model on propagation-structure information is essential for improving the prediction stability and spatial generalization capability of the NRNF.

Although the proposed approach enables continuous modeling and prediction of the UUV radiated-noise PSD within a single experimental scenario, the current model is still primarily tailored to a specific environment and does not yet generalize across multiple scenarios. Future work will explore multi-scene neural-field modeling for radiated noise, for example, by explicitly parameterizing environmental variables as conditional inputs, or by jointly representing geometric, environmental, and acoustic observations through multimodal learning. These directions are expected to further enhance the model's adaptability and generalization capability across different water bodies, boundary conditions, and acoustic propagation environments.
\begin{acknowledgments}
This work was supported by the Stable Supporting Fund of Acoustic Science and Technology Laboratory,China(Grant No. JCKYS2024604 SSJS005).
\end{acknowledgments}

\section*{AUTHOR DECLARATIONS}

\subsection*{Conflict of Interest}
The authors have no conflicts to disclose.

\section*{DATA AVAILABILITY}
The data that support the findings of this study are available from the corresponding author upon reasonable request.

\bibliography{reference}

\end{document}